\documentstyle[preprint,aps]{revtex}

\tighten

\begin{document}

\preprint{SFU HEP-152-98}

\draft

\title{Tadpole-improved SU(2) lattice gauge theory}

\author{Norman H. Shakespeare
\footnote{Email address: nshakesp@sfu.ca.}
and Howard D. Trottier
\footnote{Email address: trottier@sfu.ca.}}
\address{Department of Physics, Simon Fraser University, 
Burnaby, B.C., Canada V5A 1S6}

\date{March 1998}

\maketitle

\begin{abstract}
\noindent
A comprehensive analysis of tadpole-improved SU(2) 
lattice gauge theory is made. Simulations are done on isotropic 
and anisotropic lattices, with and without improvement. 
Two tadpole renormalization schemes are employed, one using
average plaquettes, the other using mean links in Landau gauge. 
Simulations are done with spatial lattice spacings $a_s$
in the range of about 0.1--0.4~fm. Results are presented for the 
static quark potential, the renormalized lattice anisotropy $a_t/a_s$
(where $a_t$ is the ``temporal'' lattice spacing), and for the scalar 
and tensor glueball masses. Tadpole improvement significantly reduces 
discretization errors in the static quark potential and in the scalar 
glueball mass, and results in very little renormalization of the bare 
anisotropy that is input to the action. We also find that tadpole 
improvement using mean links in Landau gauge results in smaller 
discretization errors in the scalar glueball mass (as well as in the 
static quark potential), compared to when average plaquettes are used.
The possibility is also raised that further improvement in the scalar 
glueball mass may result when the coefficients of the operators 
which correct for discretization errors in the action are computed
beyond tree level.
\end{abstract}


\section{Introduction}

Simulations of lattice Quantum Chromodynamics (QCD) have undergone 
significant changes in the past few years, with a host of new 
actions under investigation, and with the viewpoint emerging 
that simulations on coarse lattice can yield reliable results.
A major impetus for these changes was the realization that 
the large radiative corrections which occur in many quantities in 
lattice theories have a common origin, coming from cutoff effects 
due to tadpole diagrams specific to lattice actions \cite{LepMac}.

Tadpole diagrams in lattice theories are induced by the
nonlinear connection between the lattice link variables $U_\mu(x)$
and the continuum gauge fields $A_\mu(x)$,
\begin{equation}
   U_\mu(x) \equiv e^{i a g A_\mu(x)} .
\end{equation}
The cutoff dependence of ultraviolet divergent tadpole diagrams 
spoils naive power counting in the lattice spacing $a$.
Higher dimension operators that are generated by lattice actions 
generally induce large radiative corrections, their contributions 
being suppressed by powers of $\alpha_s(a)$, rather than of $a$. 
Fortunately, there is now considerable evidence that the 
effects of tadpole diagrams can largely be removed with a 
simple mean field renormalization of the links \cite{LepMac}
\begin{equation}
   U_\mu(x) \rightarrow {U_\mu(x) \over u_0} ,
\label{u0}
\end{equation}
where an operator dominated by short-distance fluctuations
is used to determine $u_0$. (Alternative approaches to the
design of improved lattice actions include the construction 
of fixed point actions \cite{FixedPoint}, and the use of 
nonperturbative renormalization conditions \cite{LuscherReview}.)

Tadpole improvement has helped to revitalize interest in the Symanzik
improvement program \cite{Symanzik}, with a number of new more 
complex actions currently under investigation, partly with the goal 
of doing precision simulations on coarse lattices (for a review see 
Ref. \cite{LepReview}). One of the earliest applications of 
tadpole improvement was in the development of the nonrelativistic QCD 
(NRQCD) action for heavy quarks \cite{NRQCDReview}. Tadpole 
improvement is now widely used in large scale simulations of
many actions \cite{CloverReview}, and plays an important role in 
current efforts to extract continuum results from simulations 
on fine lattices. 

The current interest in simulations on very coarse lattices
was stimulated by the first study of tadpole-improved gluonic
actions \cite{LepCoarse}. Simulations of the static quark
potential, and of the spin-average charmonium spectrum in NRQCD,
using an improved gluonic action on lattices with spacings as 
large as 0.4~fm, showed discretization errors of only a few 
percent. 

This has led to more recent efforts to further optimize tadpole-improved 
actions. In particular, there has been considerable work to determine 
the optimal choice of operator to use in defining the mean field 
renormalization factor $u_0$ \cite{Lepu0L,HDT,D234,newHDT,bbscaling}.
Most previous simulations have used the fourth root of the
average plaquette $u_{0,P}$ for tadpole improvement where,
in SU(N) gauge theory (on isotropic lattices) 
\begin{equation}
   u_{0,P} \equiv
       \left\langle {1 \over N} \mbox{ReTr} \, U_{\mbox{pl}} 
       \right\rangle^{1/4} .
\label{uplaq}
\end{equation}
However simulations of the static quark potential \cite{Lepu0L}, 
of the quarkonium spectrum in NRQCD \cite{HDT,newHDT,bbscaling}, 
and of a relativistic fermion action \cite{D234} have demonstrated 
that discretization errors are further reduced when the mean link in 
Landau gauge $u_{0,L}$ is used, where
\begin{equation}
   u_{0,L} \equiv
       \left\langle {1 \over N} \mbox{ReTr} \, U_\mu \right\rangle, 
       \quad \partial_\mu A_\mu = 0 .
\label{ulan} 
\end{equation}

There has also been a rapid evolution of more complex, highly 
improved actions (see e.g. Refs. \cite{LepReview,D234,LepAn,ColinAn}). 
In particular, there has been a revival of interest in actions 
defined on anisotropic lattices, where the lattice spacing in the 
``temporal'' direction $a_t$ is kept smaller than the spatial spacing
$a_s$ \cite{LepAn,ColinAn}. This enables much more efficient 
simulations of hadronic systems with large masses, for example, 
where the exponential suppression of the correlation function 
becomes prohibitive on lattices with large $a_t$. This is especially 
relevant for glueball simulations, where discretization errors may 
be acceptably small even with $a_s$ as large as 0.4~fm
\cite{ColinGlue1,ColinGlue,Peardon}, but where 
the correlation function becomes extremely noisy after only a few
time steps if a comparable $a_t$ is used.

An impediment to the use of anisotropic lattices has been the need to 
measure the renormalized anisotropy $(a_t/a_s)_{\rm phys}$
in the simulation which, in the case of the Wilson gluon action
\cite{OldAn,Klassen}, can differ appreciably from the bare 
anisotropy $\xi$ that is input to the action 
\begin{equation}
   \xi \equiv \left( {a_t \over a_s} \right)_{\rm bare} .
\label{xi}
\end{equation}
However it has recently been shown that tadpole improvement reduces
the renormalization of $\xi$ to a few percent \cite{LepAn,ColinAn}, 
which is small enough to be neglected in many applications. Recent
simulations of a variety of glueball states with a tadpole-improved
action, on anisotropic lattices with coarse $a_s$, have provided 
results that compete with much larger scale simulations on fine 
lattices \cite{ColinGlue,Peardon}.

In this paper we present a comprehensive analysis of the 
tadpole-improved SU(2) lattice gauge theory. We do simulations on 
both isotropic and anisotropic actions, with and without improvement. 
Results are obtained with two tadpole renormalization schemes,
one using average plaquettes, and the other using mean links 
in Landau gauge. We also compare with simulations of the Wilson action. 
Simulations are done with spatial spacings $a_s$ in the range of 
about 0.1--0.4~fm. Results are presented for the static quark 
potential, the renormalized anisotropy, and for the scalar and 
tensor glueball masses. Some of our results have been reviewed 
in preliminary form elsewhere \cite{LepReview,LepAn,ColinAn,Peardon}.
A full account of this work is presented here for the first time 
\cite{NSthesis}. (Simulations of the tadpole-improved SU(2)
gauge theory on isotropic lattices have also recently been reported in 
Refs. \cite{Poulis,Pennanen}.)

We note that there is a long history of simulations of SU(2)
gauge theory, motivated by the fact that this theory exhibits much 
of the physics found in SU(3) color, including linear confinement 
of static quarks and a rich glueball spectrum. Simulations of SU(2) 
color have been used to shed light on the physics of confinement, 
and to test new algorithms for use in more realistic simulations. 
The reduced computational cost of simulations in SU(2) continues 
to be exploited in, for example, recent large scale simulations of 
the static quark potential \cite{Largesu2}. 

The work we present here on tadpole-improved SU(2) gluonic actions 
is not only of interest in reproducing the features of the SU(3) 
theory, but also serves to suggest new avenues for further 
development. Of particular interest are new results presented 
here for the SU(2) scalar glueball mass, which relate to a peculiar 
feature that has previously been observed in SU(3) coarse lattice 
simulations \cite{ColinGlue}. We find that the SU(2) scalar 
glueball mass first decreases as the lattice spacing $a_s$ is 
increased, reaching a minimum at $a_s \approx 0.3$~fm; the mass 
then gradually increases with $a_s$. A similar ``dip'' in the 
SU(3) scalar glueball mass was reported in Ref. \cite{ColinGlue}.
It has been conjectured \cite{ColinGlue,Peardon} that this behavior 
may be related to the presence of a critical endpoint in a line of 
phase transitions in gluonic actions that include an adjoint 
coupling \cite{adjoint}. 

On the other hand, we find that the depth of the dip in the SU(2) 
theory with plaquette tadpole improvement is about half of 
that found with the Wilson action.
We also find that the depth of the dip is further reduced when 
mean link tadpole improvement is used (the SU(3) simulations in 
Ref. \cite{ColinGlue} were done with average plaquette tadpoles).
This suggests that the dip may be due, at least in part,
to discretization errors that are more fully removed with 
mean link tadpoles. The tensor glueball mass exhibits relatively
small scaling violations even with the Wilson action, which could
indicate that the tensor has a larger size than the scalar. 
In this connection, we note that the scalar glueball has been 
found to be much less sensitive to finite volume effects than 
the tensor \cite{Baal}. This supports the conjecture that the
scalar glueball has a smaller size than the tensor, and hence
is more sensitive to discretization corrections to the action.

We also performed simulations after making small changes to the 
coefficients of the operators which correct for discretization errors 
in the action. These results raise the possibility that the dip in
the scalar glueball mass might be eliminated once the correct 
$O(\alpha_s)$ renormalizations of the relevant operators in the 
action are included.

The rest of this paper is organized as follows.
In Sect.~\ref{SectIso} we present results from an improved SU(2) 
action on isotropic lattices. Two independent sets of simulations 
are done, using the two tadpole renormalization schemes 
discussed above (Eqs. (\ref{uplaq}) and (\ref{ulan})).
We simulate on lattices with spacings of about
$0.25$~fm and $0.40$~fm. We compare the static quark potential 
with improvement to the results of a simulation using the Wilson action.
We also present results for the scalar glueball mass. 
Details of our Wilson loop and glueball correlation functions
are also discussed in Sect.~\ref{SectIso}.

In Sect.~\ref{SectAn} we present results from our simulations
on anisotropic lattices. We run at several spatial lattice spacings
$a_s$ in the range of about 0.1~fm to 0.4~fm, where the temporal
spacing $a_t$ is kept near 0.1~fm. We run simulations for
improved actions using the two tadpole renormalization schemes,
and we also run on several lattices using the Wilson action. We 
compare bare and renormalized anisotropies, static quark potentials,
and scalar and tensor glueball masses from these actions. 
We also illustrate the sensitivity of the scalar glueball 
mass to small changes in the coefficients of the operators which
correct for discretization errors in the action.
We briefly summarize our results and outline possible directions 
for further work in Sect.~\ref{SectCon}.

\section{Isotropic Lattices}
\label{SectIso}

The isotropic tadpole-improved action for SU(2) is identical in 
form to the SU(3) action \cite{LepCoarse}, and at tree level is 
given by
\begin{equation}
   {\cal S} = - \beta \sum_{x,\,\mu>\nu}
   \left\{
   \frac{5}{3}  \frac{P_{\mu\nu}}{u_{0}^4}
 - \frac{1}{12} \frac{R_{\mu\nu}}{u_{0}^6}
 - \frac{1}{12} \frac{R_{\nu\mu}}{u_{0}^6}
   \right\} ,
\label{Siso}
\end{equation}
where $P_{\mu\nu}$ is one half the trace of the $1\times1$ Wilson 
loop in the $\mu\times\nu$~plane, and $R_{\mu\nu}$ is one half the 
trace of the $2\times1$ rectangle in the $\mu\times\nu$~plane. 
The notation used in Eq. (\ref{Siso}) differs slightly
from that used in Ref. \cite{LepCoarse}, where a factor of 
$5/3 u_0^4$ was absorbed into the definition of $\beta$. The 
notation used here follows that introduced in 
Refs. \cite{LepAn,ColinAn}. The leading discretization errors in 
this action are of $O(\alpha_s a^2)$ and $O(a^4)$.

We ran simulations at two lattice couplings using plaquette
improvement, $u_0=u_{0,P}$ (Eq. (\ref{uplaq})), and at one lattice 
coupling using mean link improvement, $u_0=u_{0,L}$ (Eq. (\ref{ulan})).
For comparison, results were also obtained for a standard Wilson 
action on a coarse lattice. The parameters of these 
four lattices are given in Table \ref{Tbetaiso}.
In the following we will use $\beta_P$ and $\beta_L$ to denote 
the lattice couplings for the improved action with plaquette 
and mean link tadpoles respectively, and $\beta_W$ for the 
Wilson action coupling.

Results for the static quark potential $V(R)$ on the lattices
with comparable spacings are shown in Fig. \ref{FisoV}. 
The potentials were measured at integer 
separations $R/a=1$--4. Measurements were also made using non-planar 
Wilson loops, corresponding to separations $R/a = \sqrt2$, $\sqrt3$,
$\sqrt5$, and $\sqrt8$. Symmetric combinations of the shortest
spatial paths connecting two lattice points were used in the non-planar
Wilson loop calculations. The data in Fig. \ref{FisoV} were obtained
by looking for a plateau in the time-dependent effective potential
\begin{equation}
   a V(R,T) = - \ln \left[ 
   { W(R,T) \over W(R,T-1) } \right] ,
\label{VRT}
\end{equation}
where $W(R,T)$ denotes the Wilson loop. An estimate of the 
systematic error in the large $T$ extrapolation is given by 
the statistical error in the second or third time step
after the onset of the plateau. 

An iterative fuzzing procedure was used to construct Wilson loop 
(and glueball) operators with a very high degree of overlap with 
the lowest-lying state. Fuzzy link variables $U^{(n)}_i(x)$ at the 
$n$th step of the iteration were obtained from a linear combination of 
the link and surrounding staples from the previous step \cite{APE}
\begin{equation}
   U^{(n)}_i(x) = U^{(n-1)}_i(x) 
                + \epsilon \sum_{j \neq \pm i} 
                U^{(n-1)}_j(x) U^{(n-1)}_i(x+\hat\jmath) 
                U_j{^{(n-1)}}^\dagger(x+\hat\imath)  ,
\label{fuzz}
\end{equation}
where $i$ and $j$ are purely spatial indices, and where the
links were normalized to $U^\dagger U = I$ after each iteration.
Wilson loops (and glueball correlators, Eq. (\ref{phi}) below)
were constructed as usual by using the fuzzy link variables in 
place of the original links. The optimal number of iterations $n$ 
and the parameter $\epsilon$ typically varied from 
$(n,\epsilon)=(10,0.04)$ on lattices with $a \approx 0.4$~fm 
to $(n,\epsilon)=(5,0.4)$ on lattices with $a \approx 0.1$~fm. 

A defect of the action Eq. (\ref{Siso}) is that the gluon 
propagator acquires a high energy pole with a negative residue
(at least in perturbation theory), due to terms in the action with 
two links in the time direction \cite{LepAn,ColinAn}. One consequence 
of these high energy ``doublers'' is that correlation functions exhibit 
exponential decay only at sufficiently large times. This is 
illustrated by an increase in the time-dependent potential $V(R,T)$ 
at small times, as shown in Fig. \ref{FisoVT}. This complicates 
the extraction of masses on coarse isotropic lattices, given the severe 
suppression of the correlation function at large times. This problem 
can be avoided by working on anisotropic lattices with sufficiently 
small $a_t$, where elimination of $O(a_t^2)$ errors in the action
may be unnecessary, as described in the next section.

The solid lines in Fig. \ref{FisoV} show the results of a fit
of the on-axis data to a standard infrared parameterization 
of the continuum potential:
\begin{equation}
   V_{\rm fit}(R) = \sigma R - {b \over R}  + c .
\label{Vfit}
\end{equation}
Infrared fluctuations of a one-dimensional string give $b=\pi/12$ 
\cite{LuscherString}. However much better fits were obtained by 
leaving $b$ as a free parameter. We typically found $b \approx 0.1$ 
on the coarsest lattices, with the string value emerging on the 
finest lattices analyzed in Sect.~\ref{SectAn} (see also 
Ref. \cite{ColinGlue}). The lattice spacings $a$ were extracted 
by matching the fit values for the string tension in lattice
units to a physical value of $\sqrt\sigma = 0.44$~GeV.

Discretization errors in the lattice action break Lorentz symmetry,
which is clearly visible in the off-axis potentials \cite{LepCoarse}.
These errors are dramatically reduced when the action is
tadpole improved, as seen in Fig. \ref{FisoV}. As a quantitative
measure of the improvement, we compare the potential
measured in the simulation from non-planar Wilson loops
with an interpolation to the on-axis data:
\begin{equation}
   \Delta V(R) \equiv 
   {V_{\rm sim}(R) - V_{\rm fit}(R) \over \sigma R} .
\label{DeltaV}
\end{equation}
Results for $R = (a,a,a)$ are given in Table~\ref{Tbetaiso}. With 
tadpole improvement the error is only a few percent even on 
lattices with spacings as large as 0.4~fm, compared to an error of 
about 30--40\% for the Wilson action \cite{LepCoarse}.
Discretization errors are further reduced when tadpole improvement
is done with $u_{0,L}$, compared to when $u_{0,P}$ is used.

Glueball correlation functions $G(T)$ were also calculated
\begin{equation}
   G(T) = \langle \phi(T) \phi(0) \rangle
        - \langle \phi \rangle^2 ,
\label{GT}
\end{equation}
where $\phi(T)$ denotes a linear combination of the trace of 
$1\times 1$ (fuzzy) plaquettes $P_{ij}(\vec x, T)$, summed over
spatial positions. We considered linear combinations which, 
in the continuum limit, excite scalar ($J^{P} = 0^{+}$) and tensor 
($J^{P} = 2^{+}$) glueballs:
\begin{eqnarray}
   & & \phi_{0^{+}}(T) = \sum_{\vec x} \left[
       P_{12}(\vec x, T) + P_{13}(\vec x, T) + P_{23}(\vec x, T)
       \right] ,
\nonumber \\
   & & \phi_{2^{+}}(T) = \sum_{\vec x} \left[
       P_{12}(\vec x, T) - P_{13}(\vec x, T) \right] ,
\label{phi}
\end{eqnarray}
together with two other linearly independent combinations 
of plaquettes for the tensor channel. Effective masses $m(T)$
were computed in the usual way:
\begin{equation}
   m(T) = -\ln \left( {G(T) \over G(T-1)} \right) .
\label{mT}
\end{equation}

An effective mass plot for the scalar glueball on the improved 
lattice with $a \approx 0.26$~fm is shown in Fig.~\ref{FisoGlueT}. 
The data are very noisy even at $T$ of two or three lattice units, 
due to the large lattice spacing and glueball mass, despite the 
fact that an ensemble of over 30,000 configurations was generated.
The data for the tensor glueball were too noisy to be of use.
As with the SU(3) simulations reported in Ref. \cite{ColinGlue1}, 
the extraction of the glueball mass from a plateau in $m(T)$ 
is computationally demanding on coarse isotropic lattices. 
As shown in the next section a far more efficient approach, 
proposed in Refs. \cite{LepAn,ColinAn}, is to simulate on anisotropic 
lattices with small spacings in the time direction.
On the other hand, the results in Fig.~\ref{FisoGlueT} compare 
favorably with Wilson action calculations on fine lattices 
\cite{WilsonData}. In Ref. \cite{Teper} a continuum extrapolation 
of Wilson action glueball data was made, with the result 
$m_{0^{+}} / \sqrt\sigma = 3.87 \pm 0.12$.

\section{Anisotropic Lattices}
\label{SectAn}

The tadpole-improved SU(2) action on anisotropic lattices
is again identical in form to the SU(3) theory and, following 
Refs. \cite{LepAn,ColinAn}, we omit corrections for $O(a_t^2)$ 
errors by working on lattices with small $a_t$. 
This has the advantage of eliminating a negative residue high energy 
pole in the gluon propagator. The resulting action has rectangles 
$R_{st}$ that extend only one lattice spacing in the time direction:
\begin{eqnarray}
   {\cal S} &=& - \beta \sum_{x,\,s>s^\prime}
   \xi \,\left\{
   \frac{5}{3}  \frac{P_{ss^\prime}}{u_s^4}
   - \frac{1}{12} \frac{R_{ss^\prime}}{u_s^6}
   - \frac{1}{12} \frac{R_{s^\prime s}}{u_s^6} \right\} 
 \nonumber \\
   && - \beta \sum_{x,\,s}
   {1 \over \xi} \,\left\{
   \frac{4}{3}  \frac{P_{st}}{u_s^2 u_t^2}
   - \frac{1}{12} \frac{R_{st}}{u_s^4 u_t^2} \right\} ,
\label{San}
\end{eqnarray}
where $\xi$ is the bare anisotropy, Eq. (\ref{xi}). ``Diagonal''
correlation functions computed from this action decrease monotonically 
with time. 

On an anisotropic lattice one has two mean fields $u_t$ and
$u_s$. A natural way to determine them is to use the mean links
in Landau gauge \cite{LepReview}
\begin{equation}
  u_t = \left\langle \case12 \mbox{ReTr} \, U_4 \right\rangle, 
  \quad
  u_s = \left\langle \case12 \mbox{ReTr} \, U_s \right\rangle, 
\label{utslan}
\end{equation}
where a lattice version of the continuum Landau gauge condition 
$\partial_\mu A_\mu = 0$ is obtained by maximizing the quantity
\begin{equation}
   \sum_{x,\mu} {1\over u_\mu a_\mu^2} \mbox{ReTr} \, U_\mu(x) .
\end{equation}
Alternatively one can define the mean fields using the measured 
values of the average plaquettes. One possibility is to first compute
$u_s$ from spatial plaquettes, $u_s^4 = P_{ss'}$, and then
to compute $u_t$ from temporal plaquettes, $u_t^2 u_s^2 = P_{st}$.
However, in the limit that $a_t/a_s \to 0$, this procedure yields
$u_t \to 1/u_s$ \cite{LepPrivate}; with mean link tadpoles 
Eq. (\ref{utslan}) on the other hand one has the more physical 
limit $u_t = 1 - O((a_t/a_s)^2)$. Since the lattice 
spacing $a_t$ in our simulations is small, we adopt the following 
prescription \cite{LepAn,ColinAn,ColinGlue} for the mean fields in
``plaquette improvement'' 
\begin{equation}
   u_t \equiv  1 ,
   \quad
   u_s = \langle P_{s s'} \rangle^{1/4} .
\label{utsplaq}
\end{equation}
In this connection, we note that the factor $u_t$ in the action 
Eq. (\ref{San}) can in fact be absorbed into a redefinition of 
$\beta$ and of the input anisotropy, according to
$\beta / u_t \to \beta$, and $\xi u_t \to \xi$.
As shown below one finds very little renormalization of the
input anisotropy $\xi$ with either of the tadpole
renormalization schemes Eqs. (\ref{utslan}) or (\ref{utsplaq}).

Simulations were performed on four lattices with plaquette improvement,
and three lattices with mean link tadpoles. In addition, we ran
simulations on six lattices with the Wilson action.
The parameters of these thirteen lattices are given in
Table~\ref{Tbetaan}. The spatial spacings $a_s$ lie in the range of
about 0.1~fm to 0.4~fm, with temporal spacings $a_t$ kept near 0.1~fm.

Configurations were generated using a
heat bath algorithm. The number of updates between measurements
varied from 10 on the coarsest lattices to 20 on the finest;
integrated autocorrelation times were computed, and satisfied
$\tau_{\rm int} \alt 0.5$ in all cases. Ensembles of about
2,000 configurations were used to measure the static quark
potential, while between 80,000 and 160,000 configurations
were generated at each $\beta$ for glueball measurements.

Tadpole improvement eliminates most of the renormalization of the 
input anisotropy $\xi$, as was first shown 
in Refs. \cite{LepAn,ColinAn}.
This can be demonstrated by comparing the static quark potential
computed from Wilson loops $W_{xt}$, where the time axis is taken 
in the direction of small lattice spacings $a_t$,
\begin{equation}
   W_{xt}(R=n_1 a_s,T=n_2 a_t) \
   \stackrel{\displaystyle\longrightarrow}{\scriptstyle n_2\to\infty} \
   Z_{xt} e^{-n_2 a_t V(R)} ,
\label{Wxt}
\end{equation}
with the potential computed from Wilson loops $W_{xy}$ with both 
axes taken in the direction of large lattice spacings $a_s$
\begin{equation}
   W_{xy}(R=n_1 a_s,T=n_2 a_s) \
   \stackrel{\displaystyle\longrightarrow}{\scriptstyle n_2\to\infty} \
   Z_{xy} e^{-n_2 a_s V(R)} .
\label{Wxy}
\end{equation}
The physical anisotropy is determined after an unphysical
constant is removed from the potentials, by subtraction of
the simulation results at two different radii
\begin{equation}
  \left({a_t \over a_s}\right)_{\rm phys} =
  { a_t V_{xt}(R_2) - a_t V_{xt}(R_1)   \over
    a_s V_{xy}(R_2) - a_s V_{xy}(R_1)  } ,
\label{xiphys}
\end{equation}
where $a_t V_{xt}$ and $a_s V_{xy}$ are the potentials in the
lattice units relevant to the two sets of Wilson loops
$W_{xt}$ and $W_{xy}$, respectively (alternative methods for
determining the renormalized anisotropy are analyzed in
Ref. \cite{Klassen}).

We computed the anisotropy twice, using two different radii $R_1$ 
for the subtraction, with fixed $R_2=2a_s$. We compare
the anisotropies determined with $R_1=a$ and $\sqrt2 a$ in
Table \ref{Taniso}. Although it is advantageous to use the
potential at smaller $R$, where the statistical errors are 
smaller, it is important to check that we are not sensitive
to possible discretization errors of $O(a_s^4/R^4)$ \cite{LepPrivate}.
The two determinations of the anisotropy are in fact in
excellent agreement.

These results show that the input anisotropy $\xi$ is renormalized
by less than a few percent when the action is tadpole improved
\cite{LepAn,ColinAn}, at least over the wide range of lattices
analyzed here. The renormalization is small enough
that one need not measure the anisotropy in many applications. 
This is to be contrasted with the Wilson action, where the 
measured value of $a_t / a_s$ is found to be about 20\% lower
than $\xi$ on the lattices analyzed here.

Sample results for the potentials both with and without tadpole 
improvement are shown in Fig.~\ref{FanV}.
The renormalization of the input anisotropy in the case of the Wilson 
action is plainly visible as a difference in slope of the potentials
computed from $W_{xt}$ and $W_{xy}$. A significant reduction
in rotational symmetry breaking with the improved action is also 
apparent in the off-axis potentials, as illustrated in Fig.~\ref{FanV}.

The scalar and tensor glueball correlators were calculated
on all thirteen lattices. Representative effective mass plots are 
shown in Fig.~\ref{FanGlueT}. The fuzzy correlators described in 
Sect.~\ref{SectIso} (Eqs. (\ref{fuzz}) and (\ref{phi})) were used.
We note that the efficiency of these calculations could be improved 
by using a variational basis of several operators, rather than just 
the fuzzy $1\times1$ plaquette used here \cite{ColinGlue}.
Our final results for the masses were obtained from single 
exponential fits to the correlation functions. In most cases
acceptable fit results for the scalar glueball were obtained 
from the fit interval $T=(3\mbox{--}5)a_t$ (we required that 
these fits be in good agreement with fits using larger intervals 
in $T$). The results are given in Table~\ref{Tglueballs}.

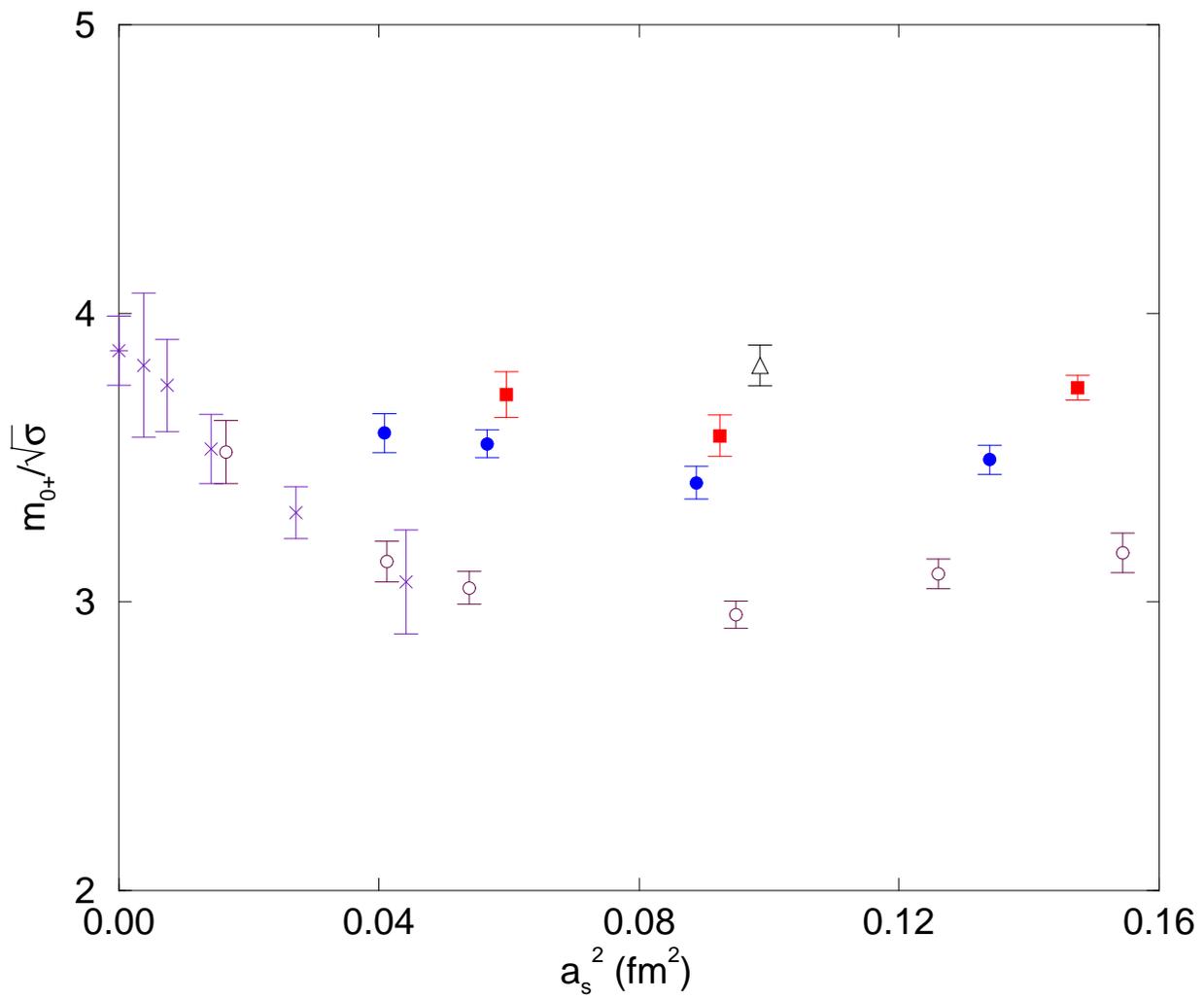
\begin{figure}
\caption{Scalar glueball mass versus spatial lattice spacing 
squared. Data is shown for:
the mean link-improved action (\protect\rule{2mm}{2mm});
the plaquette-improved action ($\bullet$); 
and from the Wilson action simulations done here ($\circ$).  
The result of a simulation with an adhoc change to the coefficients
of the rectangle terms in the improved action, described in the text, 
is also shown ($\triangle$). We also include results from Wilson 
action simulations reported in Ref. \protect\cite{WilsonData} 
($\times$). The star shows the results of an $a \to 0$ extrapolation 
of published Wilson action data \protect\cite{Teper}.}
\label{FanScalar}
\end{figure}

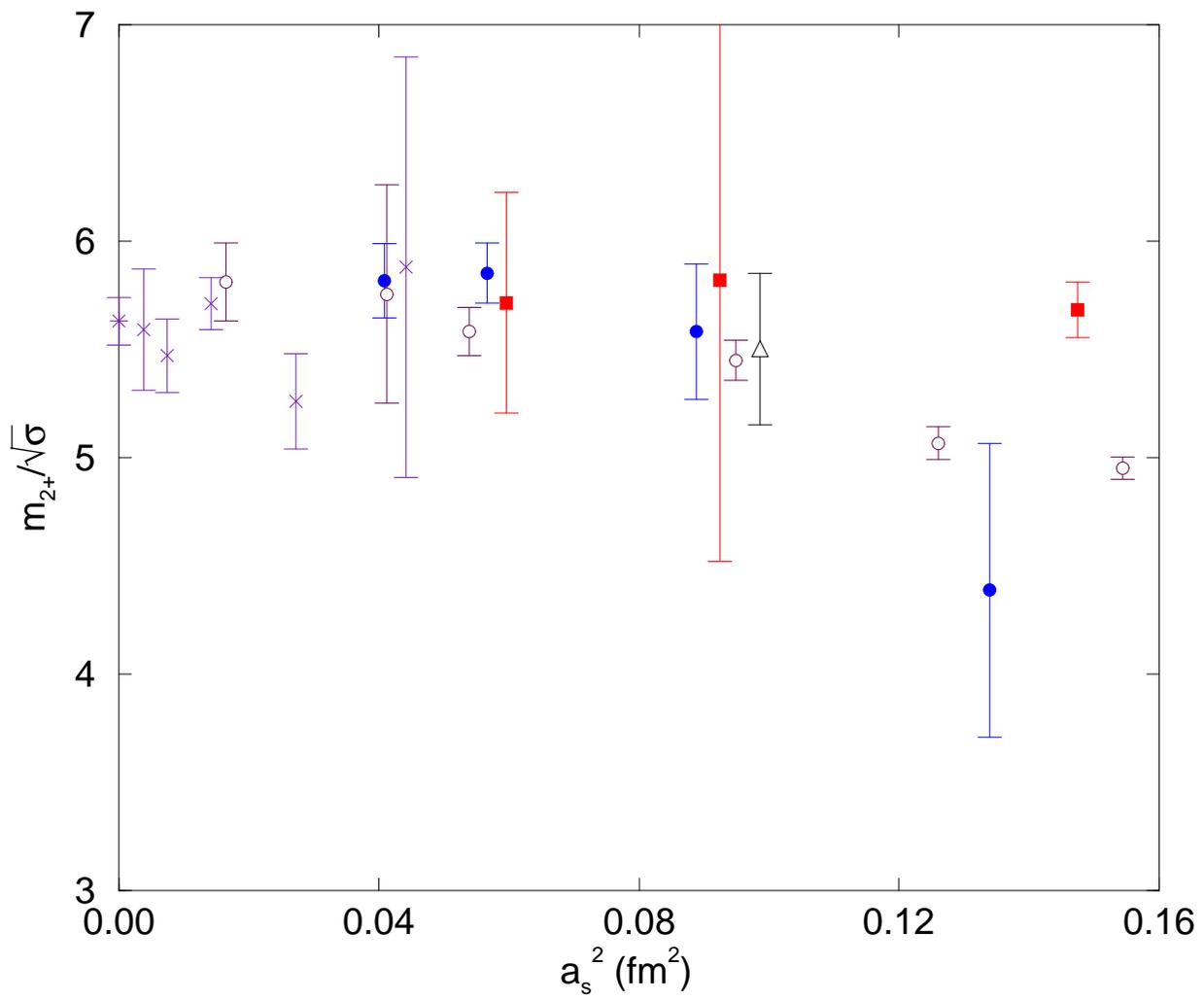
\begin{figure}
\caption{Tensor glueball mass versus lattice spacing squared. 
The plotting symbols are the same as in Fig. \ref{FanTensor}.}
\label{FanTensor}
\end{figure}

The scalar and tensor glueball masses are shown as functions of 
lattice spacing squared in Figs. \ref{FanScalar} and \ref{FanTensor}. 
The tensor glueball data exhibit relatively small scaling
violations, even for the Wilson action on the coarsest
lattices studied here. This could indicate that the tensor 
glueball has a very large size. It would be interesting to compare
with the situation for the SU(3) tensor glueball, but Wilson
action data are not available on sufficiently coarse lattices
(for a compilation see Ref. \cite{ColinGlue}).

The SU(2) scalar glueball data exhibit a peculiar feature
that has previously been observed in SU(3) coarse lattice 
simulations \cite{ColinGlue}. The scalar glueball mass first 
decreases as $a_s$ increases, reaching a minimum at 
$a_s \approx 0.3$~fm; the mass then gradually increases
with $a_s$. It has been conjectured \cite{ColinGlue,Peardon}
that this ``dip'' may be related to the presence of a critical 
endpoint in a line of phase transitions in gluonic actions that 
include an adjoint coupling \cite{adjoint}. 

The data obtained here shed new light on this behavior.
In particular, we find that the depth of the dip is reduced 
by about half when the plaquette-improved action is used,
compared to results obtained with the Wilson action. 
We also find that the depth of the dip is further reduced when 
mean link improvement is used. This suggests that the dip may 
be due, at least in part, to discretization errors that are 
more fully removed with mean link tadpoles. 

We note that the SU(3) simulations in Ref. \cite{ColinGlue} were 
done with plaquette improvement. The SU(3) scalar glueball dip is 
more pronounced than in SU(2) with plaquette improvement, but 
Wilson action data are not available on sufficiently coarse 
lattices to allow a trend to be established in 
the SU(3) theory. Simulations with mean link improvement for SU(3) 
glueballs have not been done. One the other hand, some SU(3) 
simulations have been done with an adhoc change to the action, 
designed to move the theory away from the critical endpoint in 
the fundamental-adjoint plane \cite{Peardon}; these results suggest 
that this mechanism does play a role in the SU(3) scalar glueball dip.

In this connection we performed one last set of simulations, after
making an adhoc change to the coefficient of the rectangle
terms $R_{ss'}$ and $R_{st}$ in Eq. (\ref{San}). We multiplied
these terms by a factor of $1.2$, which is consistent with an
$O(\alpha_s)$ renormalization of these operators on lattices
with spacings in the range considered here. We ran at $\beta_L=0.85$,
with $\xi=0.333$ (after retuning the Landau gauge mean link factors,
where $u_t=0.975$ and $u_s=0.758$). The lattice spacing 
$a_s = 0.314(1)$~fm is in the middle of the dip found with the 
other actions. We find glueball masses
$m_{0^+}/\sqrt\sigma = 3.82(7)$, and 
$m_{2^+}/\sqrt\sigma = 5.5(4)$ (where a measured 
renormalization of $\xi$ by about 8\% is included). These results 
are shown as an open triangle in Fig. \ref{FanScalar} and 
\ref{FanTensor}. 
These results raise the possibility that the dip in the 
scalar glueball mass might be eliminated once the correct 
$O(\alpha_s)$ renormalizations of the operators which correct for 
discretization errors in the action are included. This conjecture
is supported by finite volume studies of the glueball spectrum,
which show that the scalar is much less sensitive to finite volume
effects than the tensor \cite{Baal}.

\section{Summary and Outlook}
\label{SectCon}

Tadpole improved SU(2) lattice gauge theory was applied to 
calculations of the heavy quark potential,  the renormalized
lattice anisotropy, and the scalar and tensor glueball masses. We 
analyzed improved actions on isotropic and anisotropic lattices, and 
we compared simulations with mean link and plaquette tadpole improvement.
Comparisons were also made with simulations of the Wilson action.
Tadpole improvement significantly reduces discretization errors in the
static quark potential, and results in very little renormalization
of the input anisotropy. We also found a ``dip'' structure in
the scalar glueball mass, analyzed as a function of lattice spacing,
which has previously been observed in SU(3) simulations.
We found evidence that this dip may be related to discretization 
errors in the action that are more fully removed when tadpole 
improvement is done using the mean link in Landau gauge.
The possibility was also raised that further improvement in the scalar 
glueball mass may result when the coefficients of the operators 
which correct for discretization errors in the action are computed
beyond tree level.

Simulations with mean link improvement (as well as of the Wilson 
action) on coarse lattices in the SU(3) theory could help to 
clarify the origin of the dip in the scalar glueball mass. 
It might also be fruitful to analyze an improved SU(2) theory
with an adjoint coupling included.
A calculation of the scalar and tensor glueball sizes would 
also yield useful information, with the reduced computational cost 
of simulations in the SU(2) gauge theory providing further
incentive for continued study of this system.

\acknowledgments

We thank C. Morningstar, M. Peardon, and R.~M. Woloshyn for
helpful conversations. We are especially indebted to G.~P. Lepage 
for numerous suggestions and discussions which provided 
the impetus for much of this work.
This work was supported in part by the Natural Sciences and 
Engineering Research Council of Canada.


\begin{table}
\begin{center}
\begin{tabular}{cdccccc}
Action  &  $\beta$  
& $\langle \case13 \mbox{ReTr} \, U_{\mbox{pl}} \rangle^{1/4}$
& $\langle \case13 \mbox{ReTr} \, U_{\mu} \rangle^{1/4}$
& $a$ (fm)    & Volume   & $\Delta V(\sqrt3a)$ \\
\hline
Improved     & 0.730  & 0.844  & 0.779 & 0.394(2)  & $8^4$  & 0.04(1) \\
(Plaquette)  & 0.935  & 0.872  & 0.817 & 0.264(1)  & $8^4$  & 0.02(1) \\
\hline
Improved     & 0.550  & 0.845  & 0.779 & 0.404(2)  & $8^4$  & 0.01(1) \\
(Landau) \\
\hline
Wilson       & 1.700  & 0.802  & 0.742 & 0.402(2)  & $8^4$  & 0.32(4) \\  
\end{tabular}
\end{center}
\caption{Simulation parameters for three isotropic tadpole-improved
lattices, and one Wilson action lattice. The mean field
for plaquette improvement (second column) and mean link improvement
(third column) are given. The lattice spacings were determined
from the string tension, with $\sqrt\sigma=0.44$~GeV.
The measured errors $\Delta V$ in the off-axis potential at 
$R=(a,a,a)$ are also shown (see Eq. (\ref{DeltaV})).}
\label{Tbetaiso}
\end{table}

\begin{table}
\begin{center}
\begin{tabular}{ccccccc}
Action  
   & $\beta$  & $\xi$  & $u_t$  & $u_s$  & $a$ (fm) & Volume \\
\hline
Improved 
  & 0.848  & 0.276  &  1.  & 0.793  & 0.366(1)  & $8^3 \times 32$  \\
(Plaquette)
  & 0.981  & 0.333  &  1.  & 0.819  & 0.298(1)  & $8^3 \times 24$  \\ 
  & 1.114  & 0.409  &  1.  & 0.842  & 0.238(1)  & $8^3 \times 20$  \\
  & 1.214  & 0.500  &  1.  & 0.860  & 0.202(2)  & $10^3 \times 20$ \\
\hline
Improved
  & 0.650  & 0.276  & 0.977 & 0.718 & 0.384(2)  & $8^3 \times 32$  \\
(Landau)
  & 0.795  & 0.333  & 0.974 & 0.757 & 0.304(1)  & $8^3 \times 24$  \\
  & 0.905  & 0.409  & 0.966 & 0.785 & 0.244(1)  & $8^3 \times 20$  \\
\hline
Wilson 
  & 1.950  & 0.250  &       &       & 0.393(2)  & $8^3 \times 32$  \\
  & 2.000  & 0.250  &       &       & 0.355(2)  & $8^3 \times 32$  \\
  & 2.140  & 0.333  &       &       & 0.308(2)  & $8^3 \times 24$  \\
  & 2.243  & 0.400  &       &       & 0.232(2)  & $8^3 \times 20$  \\
  & 2.300  & 0.500  &       &       & 0.203(1)  & $10^3 \times 20$ \\
  & 2.400  & 1.     &       &       & 0.128(2)  & $12^4$ \\
\end{tabular}
\end{center}
\caption{Simulation parameters for the lattices analyzed in
Sect.~\ref{SectAn}. The bare anisotropies $\xi$ and the mean fields
$u_t$ and $u_s$ for tadpole improvement are shown, along with 
the spatial lattice spacings $a_s$ determined from 
the string tension.}
\label{Tbetaan}
\end{table}

\begin{table}
\begin{center}
\begin{tabular}{cccccc}
  &    &  & \multicolumn{2}{c}{$(a_t/a_s)_{\rm phys}$} \\
Action  
  & $\beta$  & $\xi$  & $(R_1=a_s)$ & $(R_1=\sqrt2 a_s)$  
             & $\Delta V(\sqrt3a)$  \\
\hline
Improved 
  & 0.848  & 0.276  &  0.273(4)     &     0.272(8)    & 0.08(1) \\
(Plaquette)
  & 0.981  & 0.333  &  0.335(5)     & \,\ 0.332(10)   & 0.05(1) \\
  & 1.114  & 0.409  &  0.413(3)     &     0.418(8)    & 0.03(1) \\
  & 1.214  & 0.500  &  0.502(3)     &     0.506(8)    & 0.03(1) \\
\hline
Improved
  & 0.650  & 0.276  & \,\ 0.277(11) & \,\ 0.269(22)   & 0.06(1) \\
(Landau)
  & 0.795  & 0.333  &  0.335(4)     &     0.330(9)    & 0.04(1) \\
  & 0.905  & 0.409  &  0.408(8)     & \,\ 0.402(12)   & 0.03(1) \\
\hline
Wilson 
  & 1.950  & 0.250  &  0.189(2)     &     0.183(7)    & 0.19(1) \\
  & 2.000  & 0.250  &  0.200(5)     & \,\ 0.214(13)   & 0.15(1) \\
  & 2.140  & 0.333  &  0.267(4)     &     0.271(7)    & 0.11(1) \\
  & 2.243  & 0.400  &  0.338(6)     & \,\ 0.335(16)   & 0.07(1) \\
  & 2.300  & 0.500  &  0.428(3)     &     0.419(9)    & 0.06(1) \\
  & 2.400  & 1.     &               &                 & 0.04(1) \\
\end{tabular}
\end{center}
\caption{Measured anisotropies $(a_t/a_s)_{\rm phys}$ compared
to the input anisotropies $\xi$ for the three actions.
Each anisotropy was measured twice, using two different radii
$R_1$ for the subtraction, with fixed $R_s=2a_s$ 
(see Eq. (\ref{xiphys})). The errors $\Delta V$ in the 
off-axis potential are also given.}
\label{Taniso}
\end{table}

\begin{table}
\begin{center}
\begin{tabular}{cdccc}
Action  
  & $\beta$  & $a_s$ (fm)  & $m_{0^+}/\sqrt\sigma$ 
                           & $m_{2^+}/\sqrt\sigma$ \\
\hline
Improved 
  & 0.848  & 0.366(1)  &     3.49(7)  &     4.4(7) \\ 
(Plaquette)
  & 0.981  & 0.298(1)  &     3.41(6)  &     5.6(3) \\ 
  & 1.114  & 0.238(1)  &     3.55(5)  &     5.9(1) \\ 
  & 1.214  & 0.202(2)  &     3.59(7)  &     5.8(2) \\ 
\hline
Improved
  & 0.650  & 0.384(2)  &     3.74(5)  &     5.7(1) \\ 
(Landau)
  & 0.795  & 0.304(1)  &     3.58(7)  & \,\ 5.8(13) \\ 
  & 0.905  & 0.244(1)  &     3.72(8)  &     5.7(5) \\ 
\hline
Wilson 
  & 1.950  & 0.393(2)  &     3.17(7)  &     5.0(1) \\ 
  & 2.000  & 0.355(2)  &     3.10(5)  &     5.1(1) \\ 
  & 2.140  & 0.308(2)  &     2.96(5)  &     5.4(1) \\ 
  & 2.243  & 0.232(2)  &     3.05(6)  &     5.6(1) \\ 
  & 2.300  & 0.203(1)  &     3.14(7)  &     5.8(5) \\ 
  & 2.400  & 0.128(2)  & \,\ 3.52(11) &     5.8(2) \\ 
\end{tabular}
\end{center}
\caption{Scalar and tensor glueball masses from anisotropic
lattices.}
\label{Tglueballs}
\end{table}


\begin{figure}
\caption{Static quark potentials on isotropic lattices:
(a) plaquette-improved action with $\beta_P=0.730$;
(b) mean link-improved action with $\beta_L=0.550$; and
(c) Wilson action with $\beta_W=1.700$.
Linear plus Coulomb fits to the on-axis potentials
are shown as the solid lines.}
\label{FisoV}
\end{figure}

\begin{figure}
\caption{Effective potential at $R=a$ on the 
isotropic tadpole-improved lattice with $\beta_P=0.730$.}
\label{FisoVT}
\end{figure}

\begin{figure}
\caption{Effective mass plot for the scalar glueball
on the isotropic tadpole-improved lattice with
$\beta_P=0.935$.}
\label{FisoGlueT}
\end{figure}

\begin{figure}
\caption{Static quark potentials on anisotropic lattices,
computed from Wilson loops $W_{xt}$ ($\bullet$) 
and $W_{xy}$ ($\Box$). The potentials in lattice units obtained 
from $W_{xt}$ have been rescaled by the input anisotropy,
and the potentials have been shifted by additive constants in 
order to set $a_s V(R=a_s) = 1$. 
Linear plus Coulomb fits to the $W_{xt}$ on-axis data are shown 
as the solid lines. Results are shown for
(a) the plaquette-improved action at $\beta_P=0.848$;
(b) the mean link-improved action at $\beta_L=0.650$;
(c) the Wilson action, at $\beta_W=1.950$.}
\label{FanV}
\end{figure}

\begin{figure}
\caption{Effective mass plots for (a) scalar and (b) tensor
glueballs on the anisotropic lattice with mean link improvement at
$\beta_L=0.795$. The result of a single exponential fit to
the scalar glueball is shown as the solid line, with dotted
lines showing the estimated error.}
\label{FanGlueT}
\end{figure}

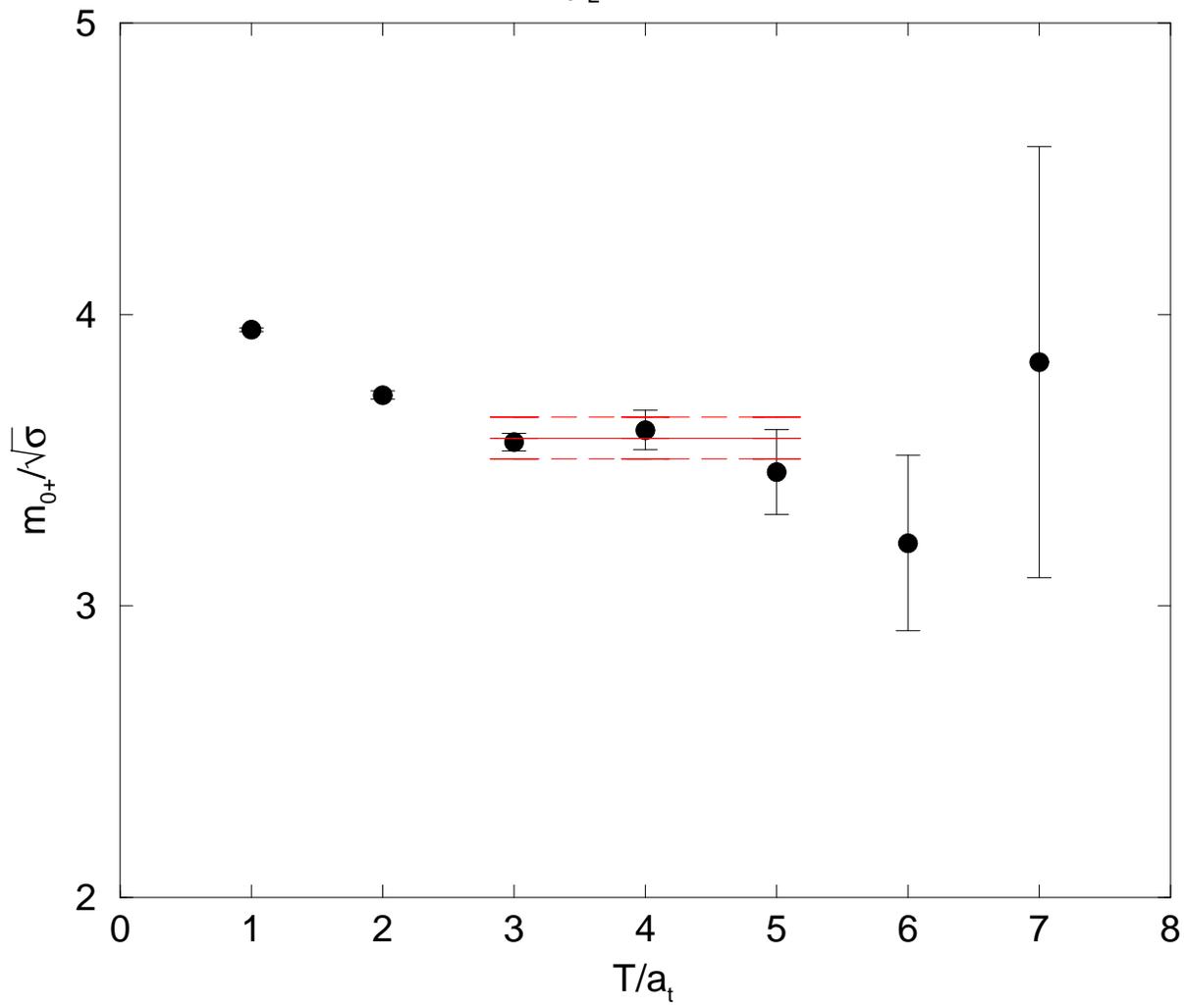
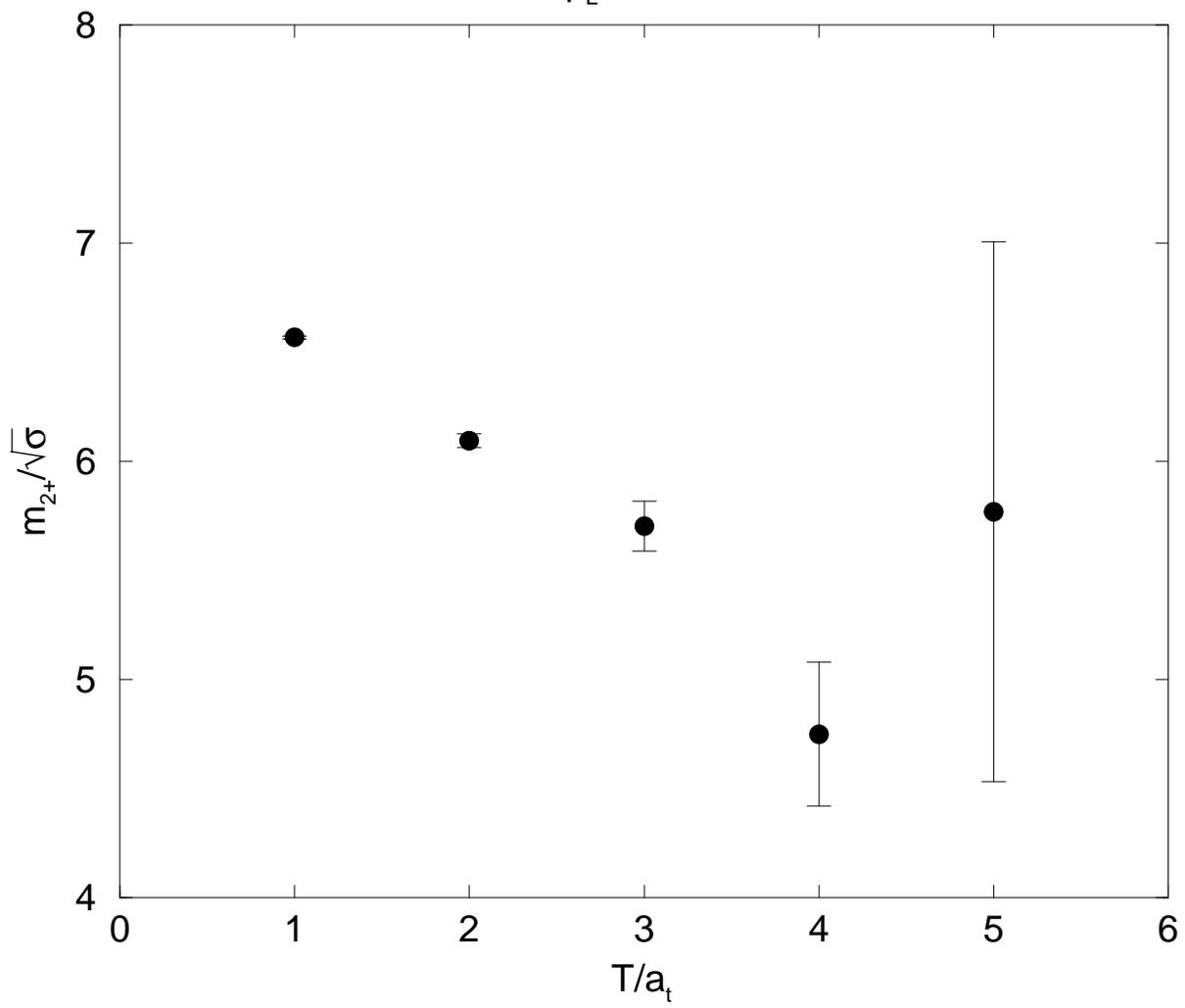
\end{document}